\documentclass[fleqn,usenatbib]{mnras}

\usepackage{newtxtext,newtxmath}

\usepackage[T1]{fontenc}

\DeclareRobustCommand{\VAN}[3]{#2}
\let\VANthebibliography\thebibliography
\def\thebibliography{\DeclareRobustCommand{\VAN}[3]{##3}\VANthebibliography}

\usepackage{graphicx}	
\usepackage{amsmath}	

\title[BRITE photometry of P Cygni]{Five years of BRITE-Constellation photometry of the luminous blue variable P Cygni: properties of the stochastic low-frequency variability\thanks{Based on data collected by the BRITE Constellation satellite mission, designed, built, launched, operated and supported by the Austrian Research Promotion Agency (FFG), the University of Vienna, the Technical University of Graz, the University of Innsbruck, the Canadian Space Agency (CSA), the University of Toronto Institute for Aerospace Studies (UTIAS), the Foundation for Polish Science \& Technology (FNiTP MNiSW), and National Science Centre (NCN).}}

\author[Elliott et al.]{Ashley Elliott,$^{1}$
Noel D. Richardson,$^{1}$
Herbert Pablo,$^{2}$
Anthony F. J. Moffat,$^{3}$
Dominic M. Bowman,$^{4}$
\newauthor Nour Ibrahim,$^{5}$
Gerald Handler,$^{6}$
Catherine Lovekin,$^{7}$
Adam Popowicz,$^{8}$ 
Nicole St-Louis,$^{3}$
\newauthor Gregg A. Wade,$^{9}$
and Konstanze Zwintz$^{10}$
\\
$^{1}$Department of Physics and Astronomy, Embry-Riddle Aeronautical University, 3700 Willow Creek Rd, Prescott, AZ, 86301, USA\\
$^{2}$American Association of Variable Star Observers, 49 Bay State Road, Cambridge, MA 02138, USA \\
$^{3}$Centre de Recherche en Astrophysique du Qu\'ebec, D\'epartement de physique, Universit\'e de Montr\'eal, Complexe des Sciences, Montr\'eal, QC H2V 0B3, Canada \\
$^{4}$Institute of Astronomy, KU Leuven, Celestijnenlaan 200D, 3001 Leuven, Belgium\\
$^{5}$Department of Astronomy, University of Michigan, 1085 S. University, Ann Arbor, MI 48109, USA \\
$^{6}$Nicolaus Copernicus Astronomical Center, Polish Academy of Sciences, Bartycka 18, PL-00-716 Warsaw, Poland \\
$^{7}$Physics Department, Mount Alison University, Sackville, NB, Canada\\
$^{8}$Department of Electronics, Electrical Engineering and Microelectronics, Silesian University of Technology, Akademicka 16, 44-100 Gilwice, Poland \\
$^{9}$Department of Physics and Space Science, Royal Military College of Canada, PO Box 17000, Station Forces, Kingston, ON, Canada K7K 7B4 \\
$^{10}$University of Innsbruck, Institute for Astro- and Particle Physics, Technikerstrasse 25/8, A-6020 Innsbruck, Austria \\
}
\date{Accepted XXX. Received YYY; in original form ZZZ}

\pubyear{2021}

\begin{document}
\label{firstpage}
\pagerange{\pageref{firstpage}--\pageref{lastpage}}
\maketitle

\begin{abstract}
Luminous Blue Variables (LBVs) are massive stars that are likely to be a transitionary phase between O stars and {hydrogen-free classical} Wolf-Rayet stars. The variability of these stars has been an area of study for both professional and amateur astronomers for more than a century. In this paper, we present five years of precision photometry of the {classical} LBV P Cygni taken with the BRITE-Constellation nanosatellites. We have analyzed these data with Fourier analysis to search for periodicities that could elucidate the drivers of variability for these stars. These data show some long-timescale variability over the course of all six calendar years of observations, but the frequencies needed to reproduce the individual light curves are not consistent from one year to the next. These results likely show that there is no periodic phenomenon present for P Cygni, meaning that the variability is largely stochastic. We interpret the data as being caused by internal gravity waves similar to those seen in other massive stars, with P Cygni exhibiting a larger amplitude and lower characteristic frequency {than the main-sequence or blue supergiant stars previously studied}. These results show evidence that LBVs may be an extrapolation of the blue supergiants, which have {previously been shown to be an extension of main-sequence stars in the context of the stochastic low-frequency photometric variability.}
\end{abstract}

\begin{keywords}
stars: massive --
stars: mass-loss --
stars: variables: S Doradus --
stars: winds, outflows
\end{keywords}

\section{Introduction}

{Luminous blue variable stars (LBVs) are massive, post main-sequence stars that have strong winds and exhibit multiple types of variability, with the time-scales of the variations ranging from days to centuries. These stars are important for understanding the evolution of stellar feedback for future star formation, despite there only being $\sim$60 known LBVs or candidate LBVs in the current census of the stars \citep{2018RNAAS...2..121R}. LBVs have been observed to have their brightness increase and exhibit supernova-like changes in their light-curves, at which time the star expels upwards of a few solar masses of material in a short time frame \citep[e.g.,][]{1994PASP..106.1025H}. The most famous of these eruptive events are those associated with the Galactic LBVs $\eta$ Carinae and P Cygni. \citet{2001A&A...366..508V} describes the properties a star must meet in order to be considered an LBV, or S Dor variable. These include visible ejecta, spectroscopic characteristics that indicate the unique characteristics of an LBV, and photometric variability that has a time scale of anywhere between days and even centuries. They also define the S Doradus phase (SD-phase) to be the time periods when the star exhibits a cyclic pattern of brightening in the optical and apparent changes in the stellar wind properties, while the star maintains a roughly constant bolometric luminosity \citep[e.g.,][]{1994PASP..106.1025H}}.

The main cause of the SD-phases, according to \citet{2001A&A...366..508V}, are the radius and the temperature variations of the star. These phases have roughly constant bolometric luminosity but the temperature and radius vary throughout these phases. The effective temperatures of these stars tend to be in the range of 15,000 to 30,000 K. There are two main types of an SD-phase, one on a timescale of years and one on a timescale of decades, referred to as the short (S) SD-phase and long (L) SD-phase, respectively. During the SD-phases, the observed magnitude of an LBV varies (up to $\sim$1-2 mag) over the span of several years or decades, due to the underlying changes in the effective temperature and radius. During the SD-phase, LBVs appear to change between a hot visual minimum state to a cool visual maximum state, during which the stars  maintain an almost constant bolometric luminosity near the Eddington limit \citep{1994PASP..106.1025H}. 

Despite several models having been developed to explain the LBV phenomenon in the context of single star evolution \citep{Grassitelli2021a}, the causes of the LBV phenomenon and the evolutionary status of these objects are both still debated, especially given the strong evidence that binary evolution dominates the evolution of massive stars \citep{2012Sci...337..444S}. In recent years, \citet{2015MNRAS.447..598S} suggested that the LBVs are actually the result of binary evolution. This hypothesis relied on LBVs being relatively isolated compared to other massive stars such as O stars and WR stars. \citet{2016ApJ...825...64H} and \citet{2018AJ....156..294A} reexamined the population of LBVs and found that the spatial distribution of other massive, hot stars was indeed similar to that of LBVs, while \citet{2019MNRAS.489.4378S} found instead that the stars are more isolated than typical O stars. Binary evolution may still allow us to explain several phenomena related to the eruptions and LBV properties. For example, the eruption of $\eta$ Carinae could be explained by a merger in a triple star system that resulted in the present-day eccentric binary \citep[e.g.,][]{2021MNRAS.503.4276H}. Furthermore, the LBVs HR Carinae and AG Carinae have been shown to be rapid rotators in certain portions of the SD-phases, namely when the star is hottest and has the smallest radius, which could be a result of previous binary mergers or mass transfer \citep{2009ApJ...705L..25G, 2009ApJ...698.1698G}. Also, some evidence exists for rapid rotation of $\eta$ Carinae \citep{2012ApJ...759L...2G}.

One of the classical LBVs with a long observational record is P Cygni. In August 1600, it made its first documented ``appearance" in the sky, reaching third magnitude and then fading below naked-eye visibility. This was the first eruption of P Cygni, discovered by Willem Janszoon Blaeu, a former student of Tycho Brahe's, who recorded the discovery in his celestial globe made in his Amsterdam workshop in 1602. A second eruption was observed in 1654. Since this time, P Cygni has become somewhat stable near fifth magnitude \citep{1994PASP..106.1025H} but also showing a gradual brightening of the star occurring over very long time scales. The long-term brightening was most recently measured to be 0.17$\pm$0.01 mag century$^{-1}$, similar to other studies of this phenomenon \citep{2011AJ....141..120R}. 

P Cygni is an ideal test bed for our understanding of LBVs, due to its enhanced brightness and relatively small distance. Spectral modeling with the non-LTE radiative transfer code CMFGEN has revealed that the star has a mass-loss rate of $3.0 \times 10^{-5} M_\odot {\rm yr}^{-1}$, a terminal wind speed of $v_\infty = 185$ km\,s$^{-1}$, and an observed effective temperature of $18,500$ K \citep{1997A&A...326.1117N}. These spectral models are derived assuming a spherically symmetric wind, which has been tested with multiple observational techniques. With long-baseline optical interferometry with the Naval Precision Optical Interferometer, \citet{2010AJ....139.2269B} found the H$\alpha$ emitting region to be spherically symmetric and stable with about a 10\% deviation from observations taken during the time period of 2005--2008. Similarly, \citet{2013ApJ...769..118R} created the first images of the wind using observations in the $H-$band with the CHARA interferometer and found that the wind was very close to spherically symmetric. \citet{2011AJ....141..120R} examined the long-term variability properties of the H$\alpha$ profile, which revealed subtle structures in the absorption trough that could be explained under the assumptions of spherical symmetry. \citet{2020ApJ...900..162G} examined over a decade of spectropolarimetry to reveal and confirm that P Cygni's wind produces intrinsic polarization almost certainly from clumping. However, the lack of preferred direction implies that the wind is spherically symmetric on the whole, in agreement with other studies \citep[e.g., ][]{1991AJ....102.1197T} since changes in the position angle happen much faster than the typical flow time of the stellar wind. 

 With the brightness and eruptive history of P Cygni, many studies of the star's variability have been carried out in an attempt to understand the properties and evolutionary status of this object. Photometric studies began in earnest with \citet{1983PASP...95..491P}, \citet{1988A&A...191..248P}, and \citet{1990ASPC....7..165D}. 
 These studies displayed three major timescales: a short period around 17 days associated with typical $\alpha$ Cygni varibility, a $\sim$100 day period similar to other known LBVs, and a cycle of years that can be classified as a short SD-phase. These timescales were confirmed by \citet{2001ASPC..233...15D}. Spectroscopically, the star has been studied most notably by \citet{2001A&A...376..898M} and \citet{2011AJ....141..120R}. \citet{2001A&A...376..898M} compared previously noted $UBV$ photometric data and new H$\alpha$ spectroscopy. These photometric data confirm the presence of a slow variation in brightness on a timescale of 7.4 years. The H$\alpha$ equivalent width determinations indicate the presence of a slow component, dubbed the Very-Long Term Component, in the variability of H$\alpha$, and is also a part of the variable SD-phase of the star. In recent years, the star has become a popular object for amateur astronomers to collect simultaneous spectroscopy and photometry, which has led to a possible detection of a {318} d period in the H$\alpha$ profile \citep{2012JAVSO..40..894P, 2013JAVSO..41...24P, 2016BAVJ....6....1P, 2020JAVSO..48..133P}.

{The spherically symmetric wind of this LBV makes the interpretation of its variability simpler to interpret. When combined with P Cygni's rich observational history and brightness, P Cygni is an ideal star with which to study LBVs.} We have analyzed five years of precision photometry from the {\it Bright Target Explorer}-Constellation nanosatellites ({\it BRITE}). We present these observations in Section 2, and our Fourier analysis in Section 3. We then discuss these observations in Section 4 and conclude our study in Section 5. 

\section{Observations}

\begin{table*}
    \centering
    \begin{tabular}{ccccc}
      Year   &  Start HJD    & End HJD   & Number of Binned Data Points  & Typical Error  \\
      \hline
      2014  &  2456847.232 &  2456924.455   &  657   &   0.00214  \\

      2015  &   2457184.674  &  2457340.628   & 2111  &  0.00172   \\

      2016  &  2457512.732  &  2457651.967    & 1731  &  0.00193    \\

      2018  &   2458274.668     &   2458385.975     & 1170  &  0.00208   \\ 
     
      2019  &   2458648.635     &   2458788.511     & 1737  &  0.00865   \\
  \hline
    \end{tabular}
    \caption{Properties of the final {\it BRITE-Toronto} light curves. }
    \label{tab:my_label}
\end{table*}

{\it BRITE-Constellation} refers to five operational nanosatellites: {\it BRITE-Austria} ({\it BAb}), {\it UniBRITE}
({\it UBr}), {\it BRITE-Lem} ({\it BLb}), {\it BRITE-Heweliusz} ({\it BHr}), and
{\it BRITE-Toronto} ({\it BTr}). The small letter appended to each
abbreviation indicating the passband in which each satellite
operates (``b" for a blue filter covering 3900--4600\AA, and ``r"
for a red filter covering 5450$-$6950\AA). The satellites are located in low-Earth
orbits with orbital periods of the order of 100 min. Each
of the $20\times20\times20$ cm$^3$ BRITE nanosatellites is equipped
with a 3-cm telescope feeding an uncooled 4008$\times$2672-pixel
KAI-11002M CCD, with a large effective unvignetted field
of view of 24$^\circ$ × 20$^\circ$
to fulfil a single purpose: tracking the
long-term photometric variability of bright stars (V$\lesssim$6) in
two passbands, typically over time periods of 2--6 months. Full technical
descriptions of the mission were provided by \citet{2014PASP..126..573W} and \citet{2016PASP..128l5001P}.

P Cygni was observed during five observation campaigns of {\it BRITE}. Our data spanned the time period from 2014 to 2019, with no observations made in 2017. For consistency, we are only using the red-filtered data, as the blue data were less consistent and much more sparse. The only {\it BRITE} satellite used for this study is {\it BRITE Toronto}. There were some data taken with the blue-filtered satellites, which were of much lower precision due to the efficiency of the system and narrower bandpass. Furthermore, the data taken with {\it UniBRITE} was found to be of too low precision in comparison to the data from the {\it BRITE Toronto} satellite. The data were processed in the standard manner for the {\it BRITE} satellites and binned to orbital means as no variability has been observed for the star with time-scales faster than 100 minutes in the past \citep{2016PASP..128l5001P,Popowicz2016ImagePI,Popowicz2017BRITEConstellationDP}. These reduced photometric data were examined and we found large jumps in relative fluxes between different portions of the light curve that were reduced separately during the pipeline process in order to mitigate problems with dark current \citep[e.g.,][]{Popowicz2018AnalysisOD, Popowicz2020MetastableDC}. To rectify this problem, we calculated linear fits near the ends of each data set in order to minimize the gap between the setups for each data set. Once a linear fit was found, the y-intercepts were subtracted from one another and that value was added to the flux data for one set so that these data were overlapping, and repeated for each small time string. This process was compared to the AAVSO $V$-band light curves to ensure the light curves provided consistent variability with the observational record. {The AAVSO light curve suffers from much lower precision than the BRITE light curves, so if the AAVSO data were too sparse in the regions where we performed a linear fit, we would compare the average flux levels of the adjacent time-series to ensure that the derived fit was reasonable. This is illustrated in Fig.~\ref{fig:reduce}, with a comparative light curve of the AAVSO data and the BRITE data shown in the supplementary material}. {In general, this process is well-supported with the comparison to the AAVSO fluxes (e.g., Fig.~\ref{fig:reduce} and Fig.~\ref{fig:appendixcompare}). There may be some additional long-period ($t\sim 3--4$ yr) noise added in our final year of data from {\it BRITE}, but the noise levels on our data are also the highest in these data.}

\begin{figure}
	\includegraphics[width=0.95\columnwidth]{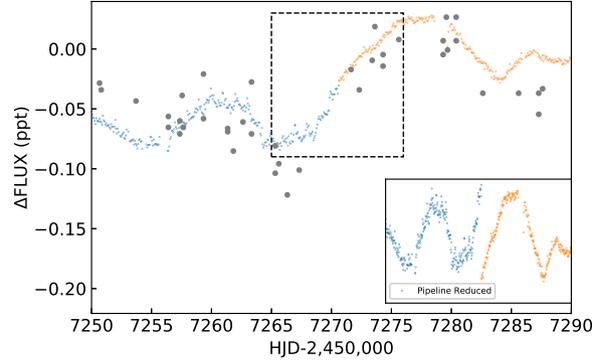}
    \caption{An example of the merging of two subsets of data from 2015 that were combined to create a more continuous data set. A linear fit was done to both the blue and orange points within the central box {with dashed lines}, and then they are shifted to match. The inset panel on the bottom right shows the pipeline-reduced and binned light curve at the same time frame, {highlighting the need for the changes}. The {final reduced data} were compared to the AAVSO $V$-band light curve to ensure consistency, {which is shown as grey points after being converted to flux here.}.  }
    \label{fig:reduce}
\end{figure}

\begin{figure}
	\includegraphics[width=\columnwidth]{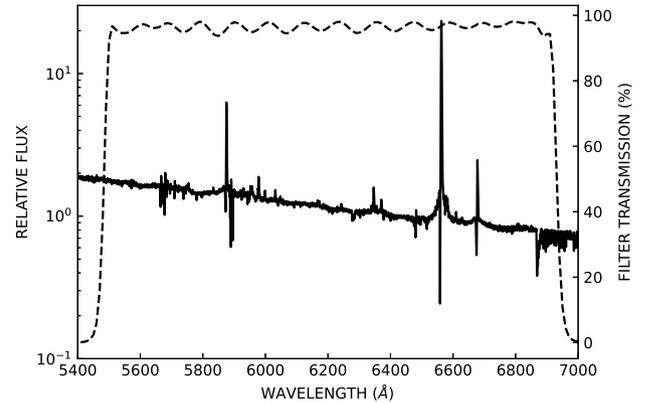}
    \caption{A typical spectrum of P Cygni, downloaded from the ARAS database, with the red {\it BRITE} filter response overplotted. }
    \label{fig:filter}
\end{figure}

Once the data were reduced into one continuous data set that showed a consistent light curve, we used AAVSO photometry to calculate average magnitudes for each year. This average magnitude was then converted into flux and subtracted from each year in order to center the data around 0. 
Given the large flux from the wind, especially in the H$\alpha$ line, we show a plot with a typical spectrum of P Cygni from the ARAS database \footnote{http://www.astrosurf.com/aras/Aras\_DataBase/LBV/PCyg.htm}
in Fig.~\ref{fig:filter} with the {\it BRITE} filter response overplotted for understanding our measurements. 

\begin{figure*}
	\includegraphics[width=\columnwidth]{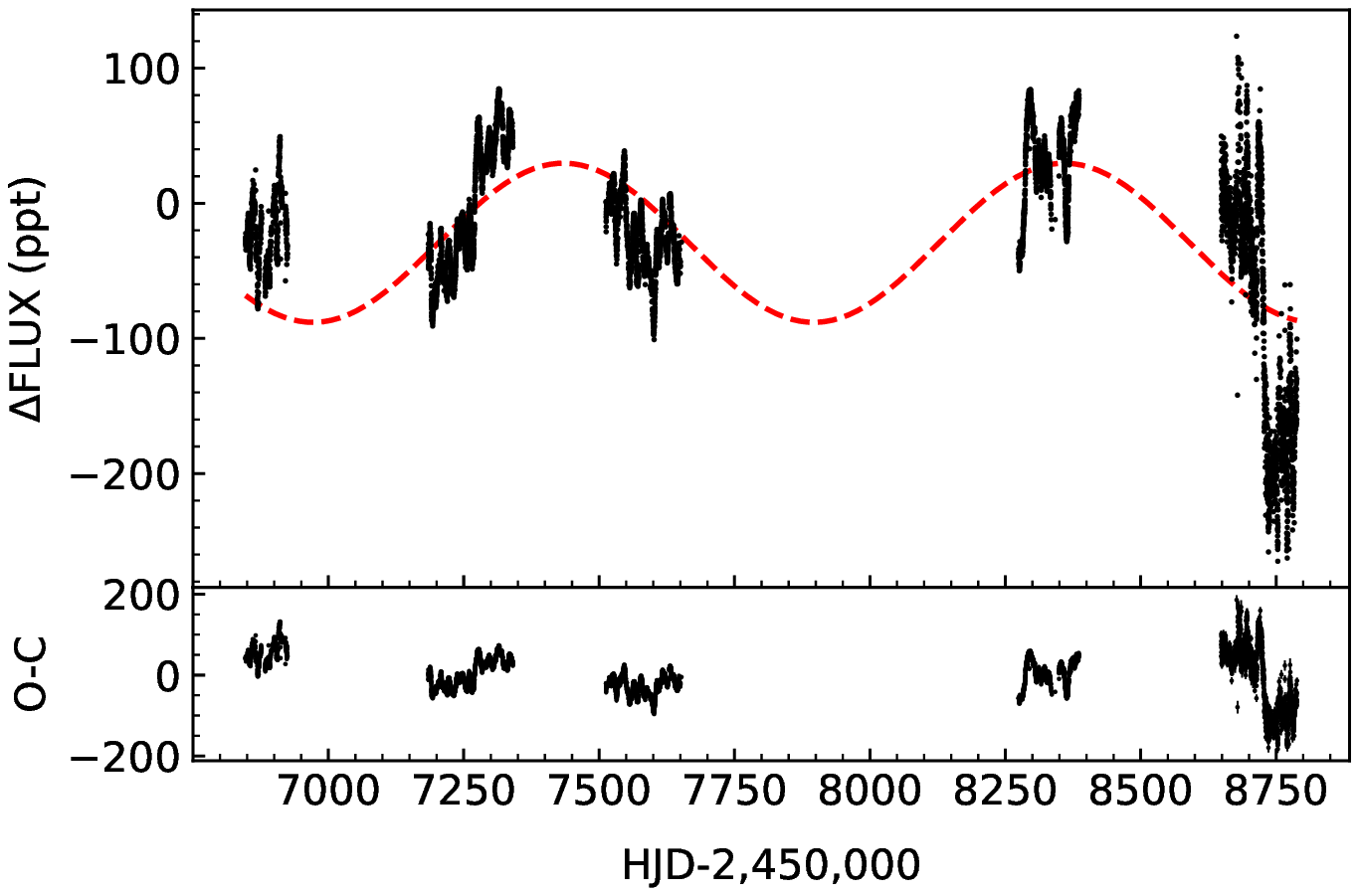}
    \includegraphics[width =\columnwidth]{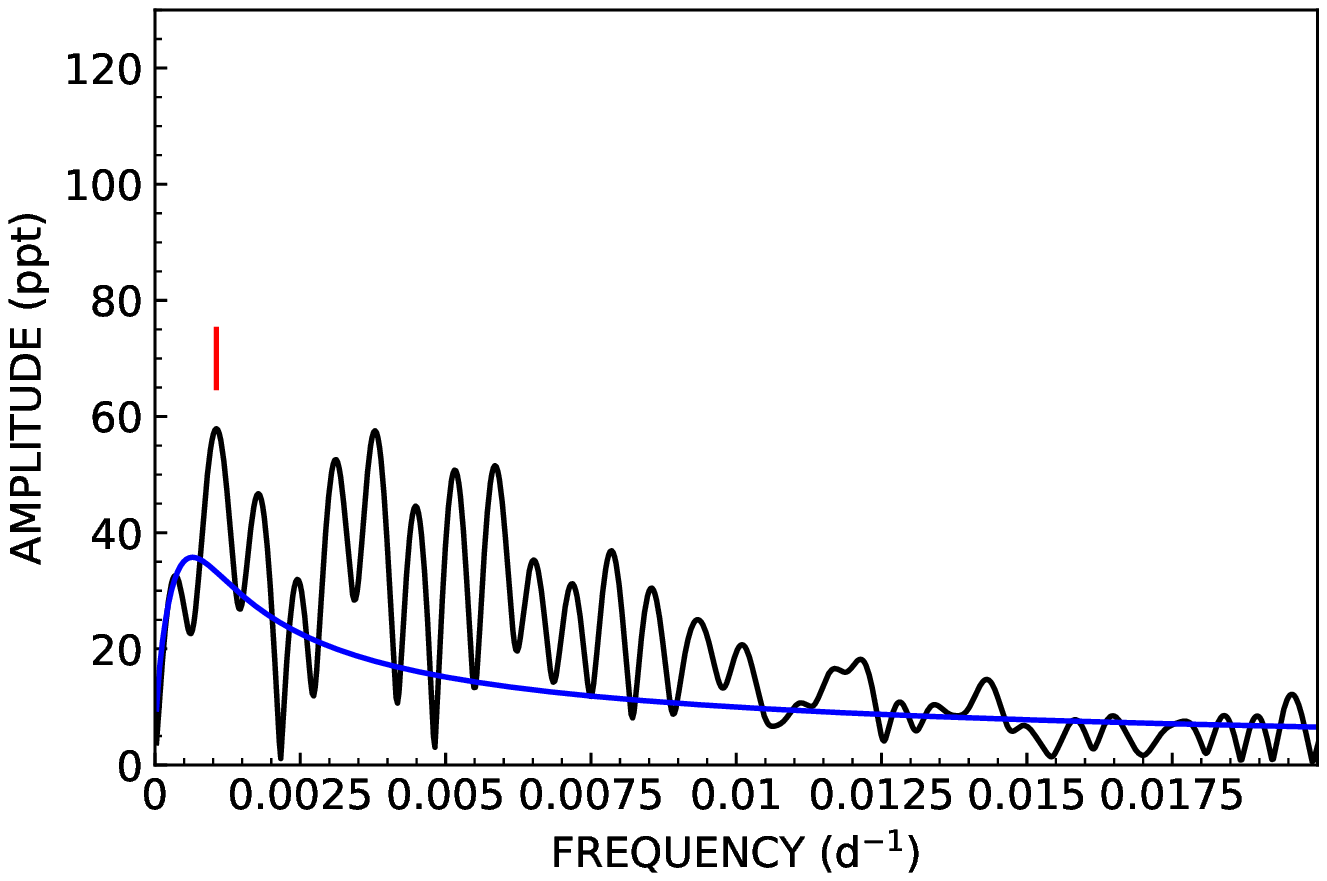}
    \caption{ In the left plot, we show all available {\it BRITE} fluxes used in our analysis, after subtracting off the global mean, with units of parts per thousand (ppt), with our calculated Fourier spectrum on the right. The frequency indicated in red on the Fourier amplitude spectrum indicates the strongest frequency from the Fourier analysis, with the blue curve representing the noise in this spectrum. The bottom panel of the photometric light curve shows the difference between the observed and calculated light curve for these data.}
    \label{fig: all data}
\end{figure*}

\begin{figure*}
	\includegraphics[width=\columnwidth]{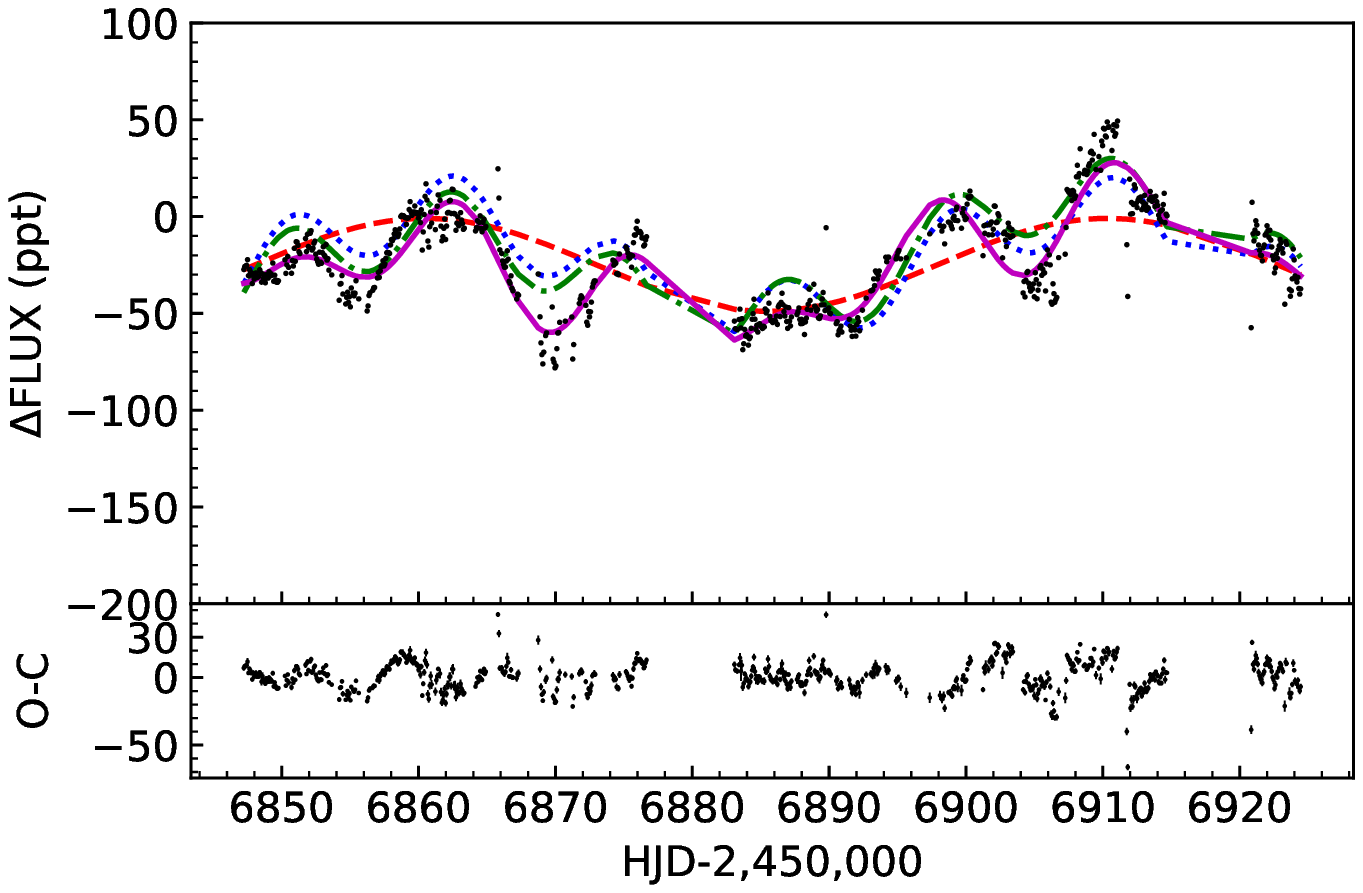}
    \includegraphics[width=\columnwidth]{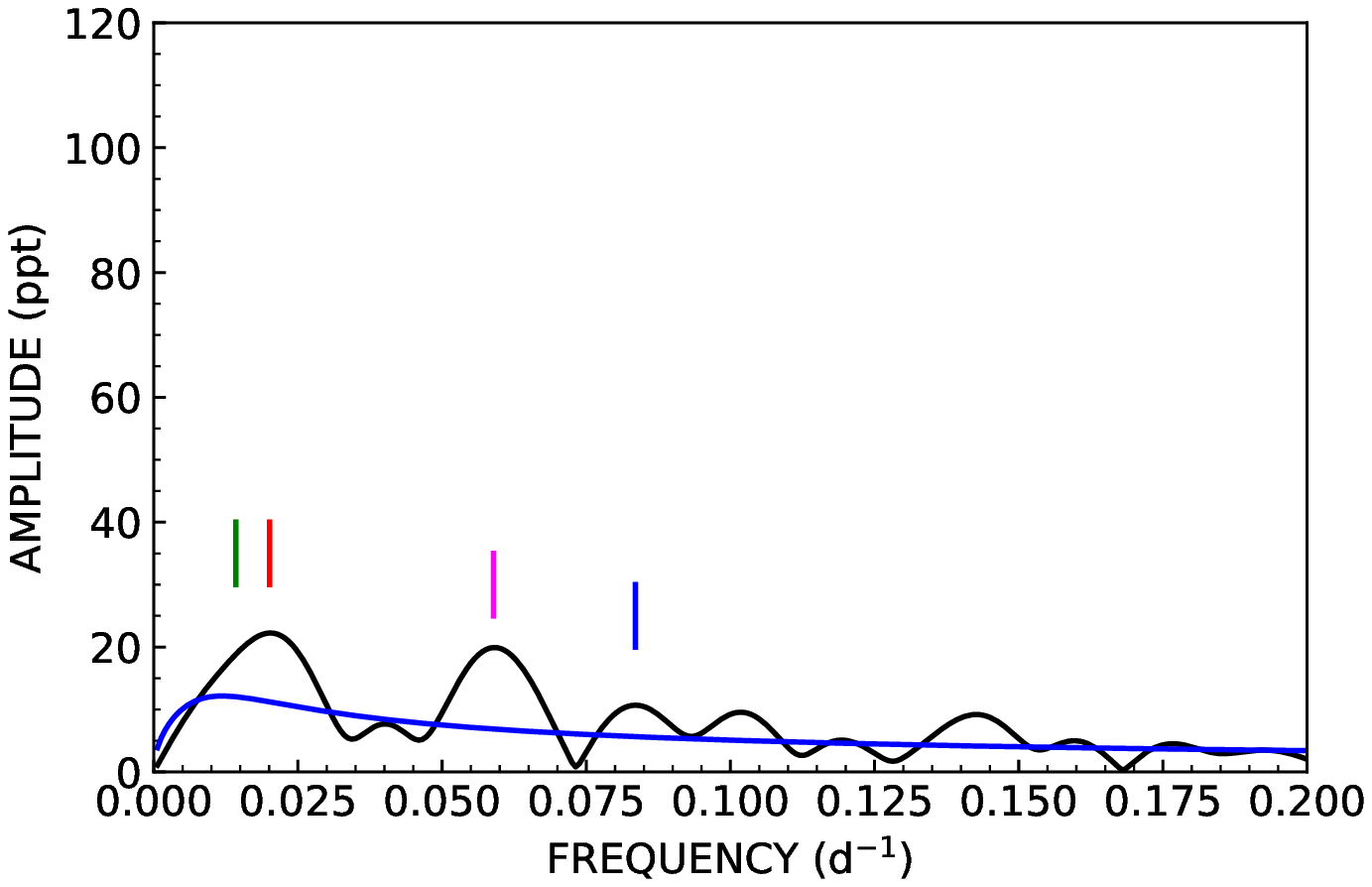}
    \caption{The {\it BRITE} flux, after subtracting off the global mean, with units of parts per thousand (ppt) (left) and the Fourier amplitude spectrum (right) for the 2014 data from {\it BRITE}. Each peak used in our analysis is highlighted with a different color in the Fourier spectrum, and then the fit is overplotted on the photometry with the corresponding color for that term and all previous terms. The final four-frequency fit is then used to calculate the $(O-C)$ that is shown on the bottom panel of the photometry. The term numbers from Table \ref{FTTable} are given by: Term 1 is shown in the red dashed line, Term 2 is shown in the dotted blue line, Term 3 is shown in the dashed and dotted green line, and Term 4 is shown as the solid pink line, where each term includes the previous ones in the fit. }
    \label{fig:2014}
\end{figure*}

\begin{figure*}
	\includegraphics[width=\columnwidth]{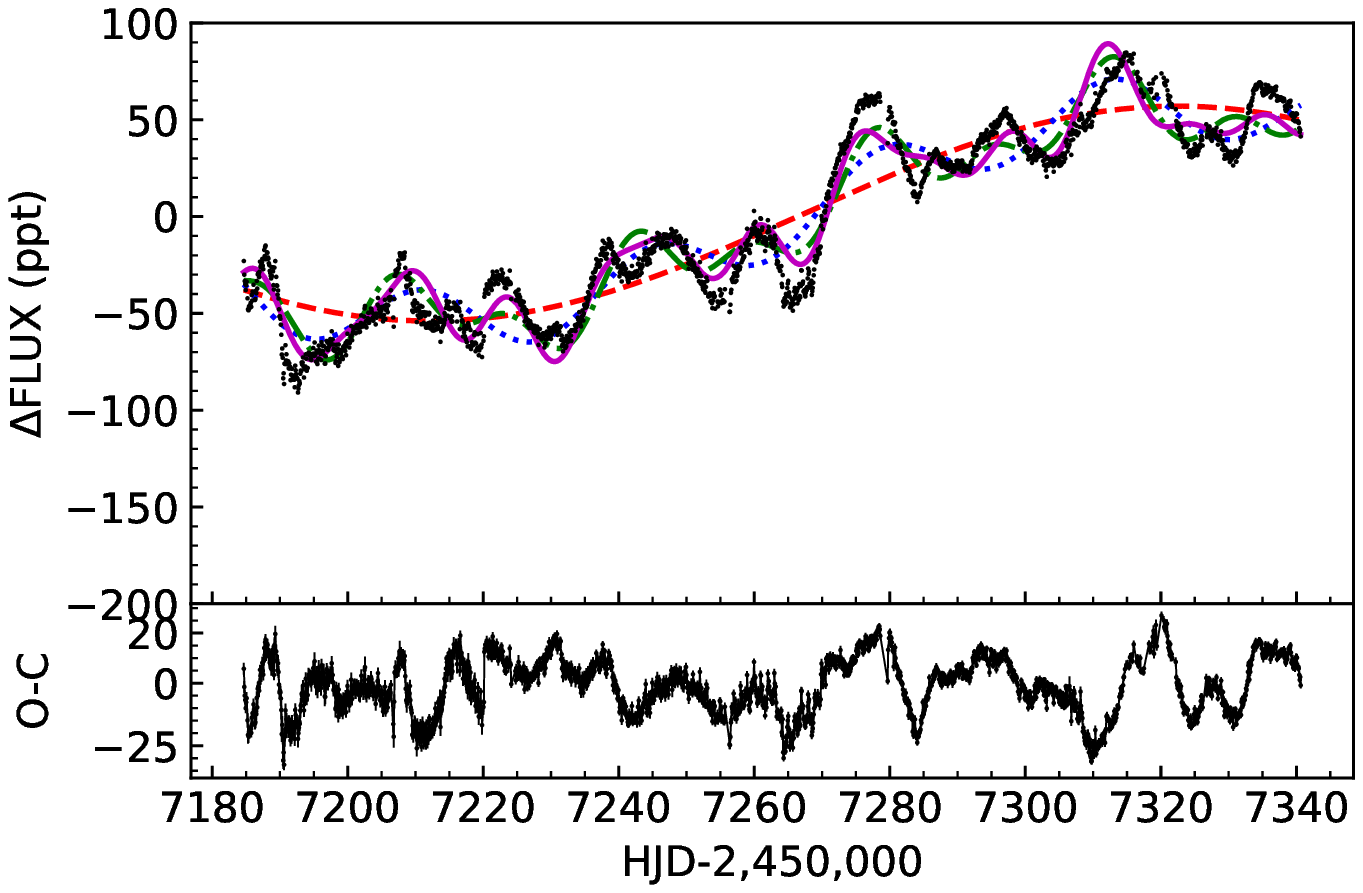}
    \includegraphics[width =\columnwidth]{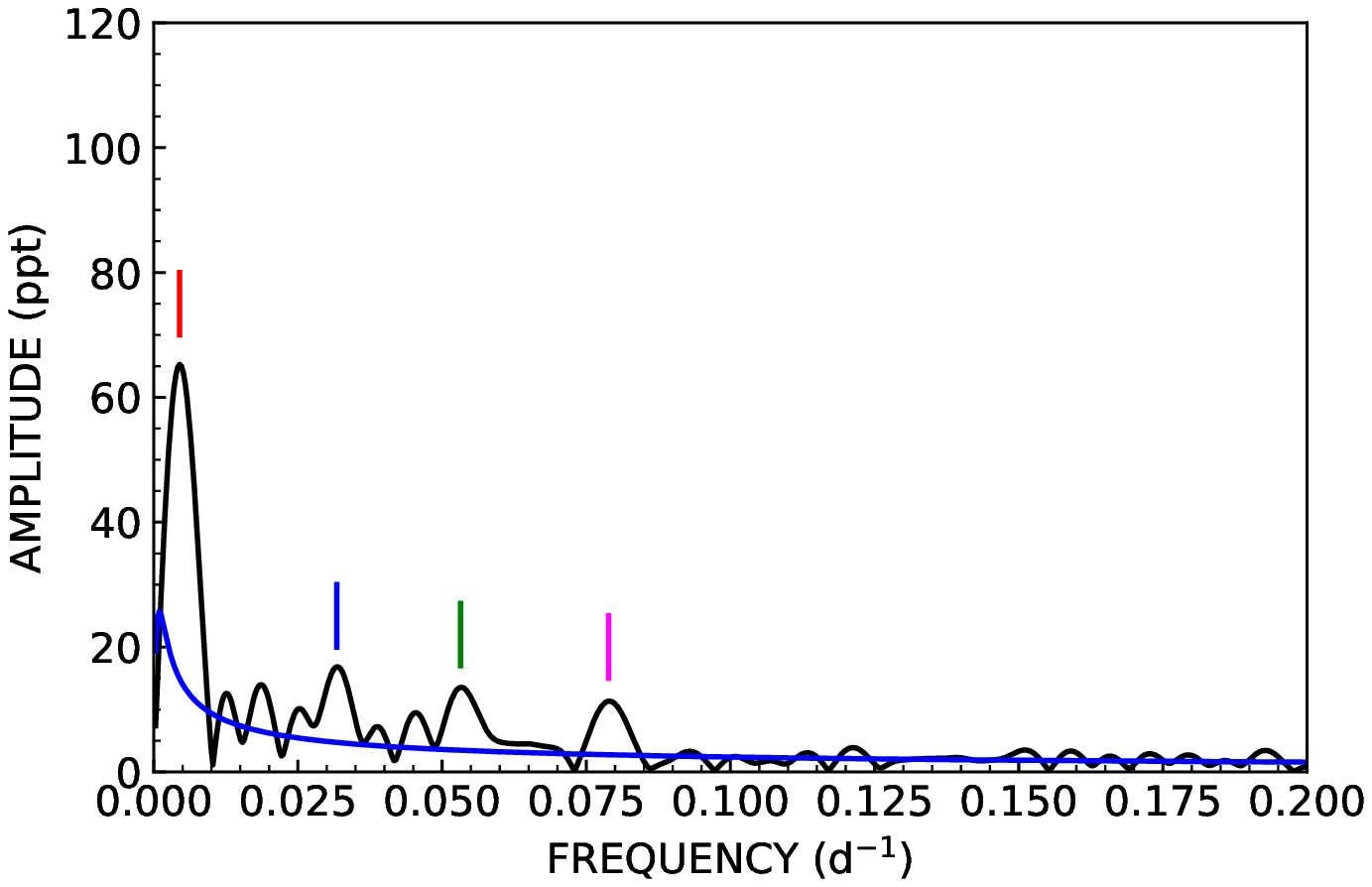}
    \caption{The {\it BRITE} flux, after subtracting off the global mean, with units of parts per thousand (ppt) (left) and the Fourier amplitude spectrum (right) for the 2015 data from {\it BRITE}, format as in Fig. \ref{fig:2014}. }
    \label{fig:2015}
\end{figure*}

\begin{figure*}
    \includegraphics[width=\columnwidth]{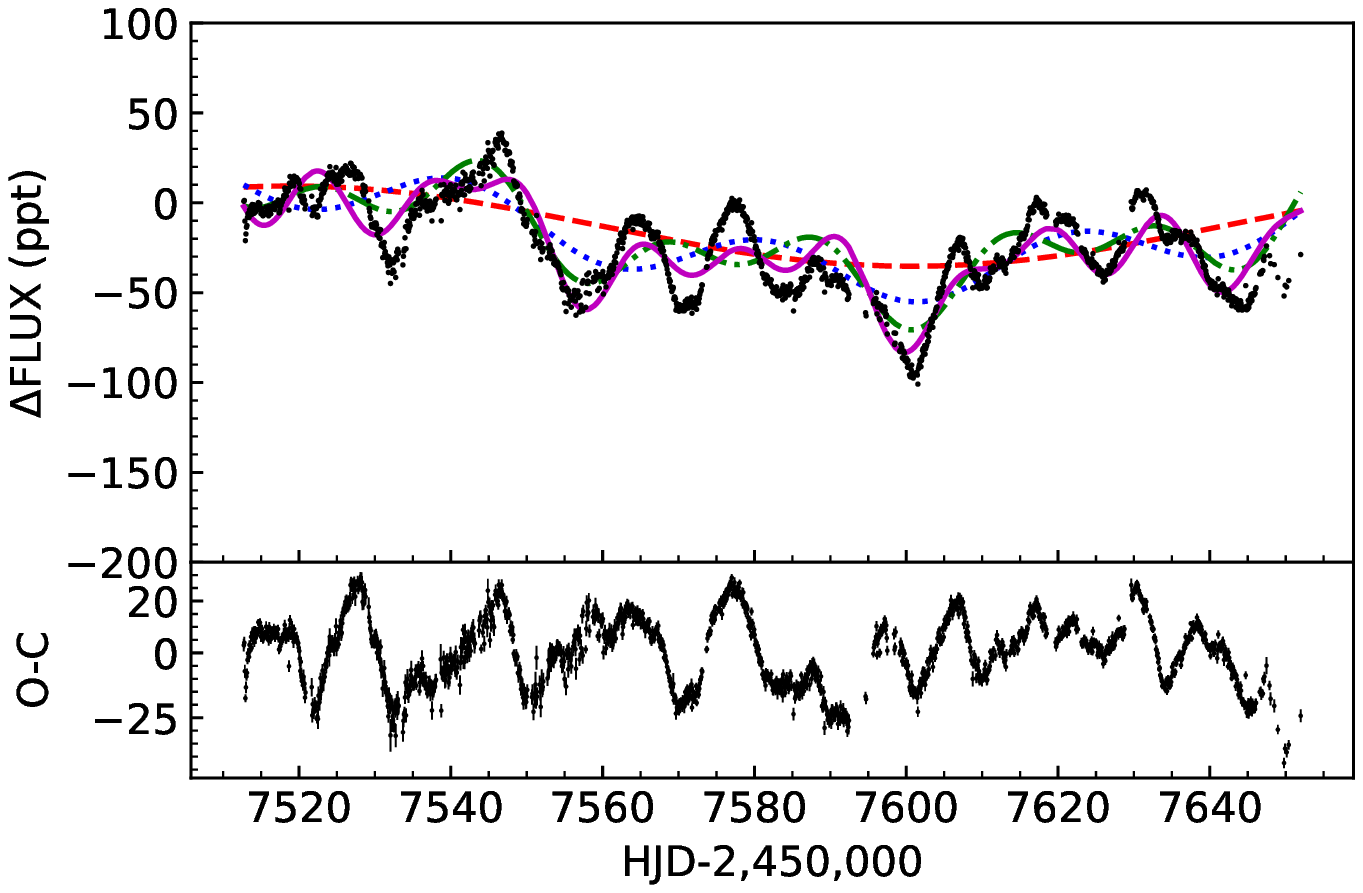}
    \includegraphics[width = \columnwidth]{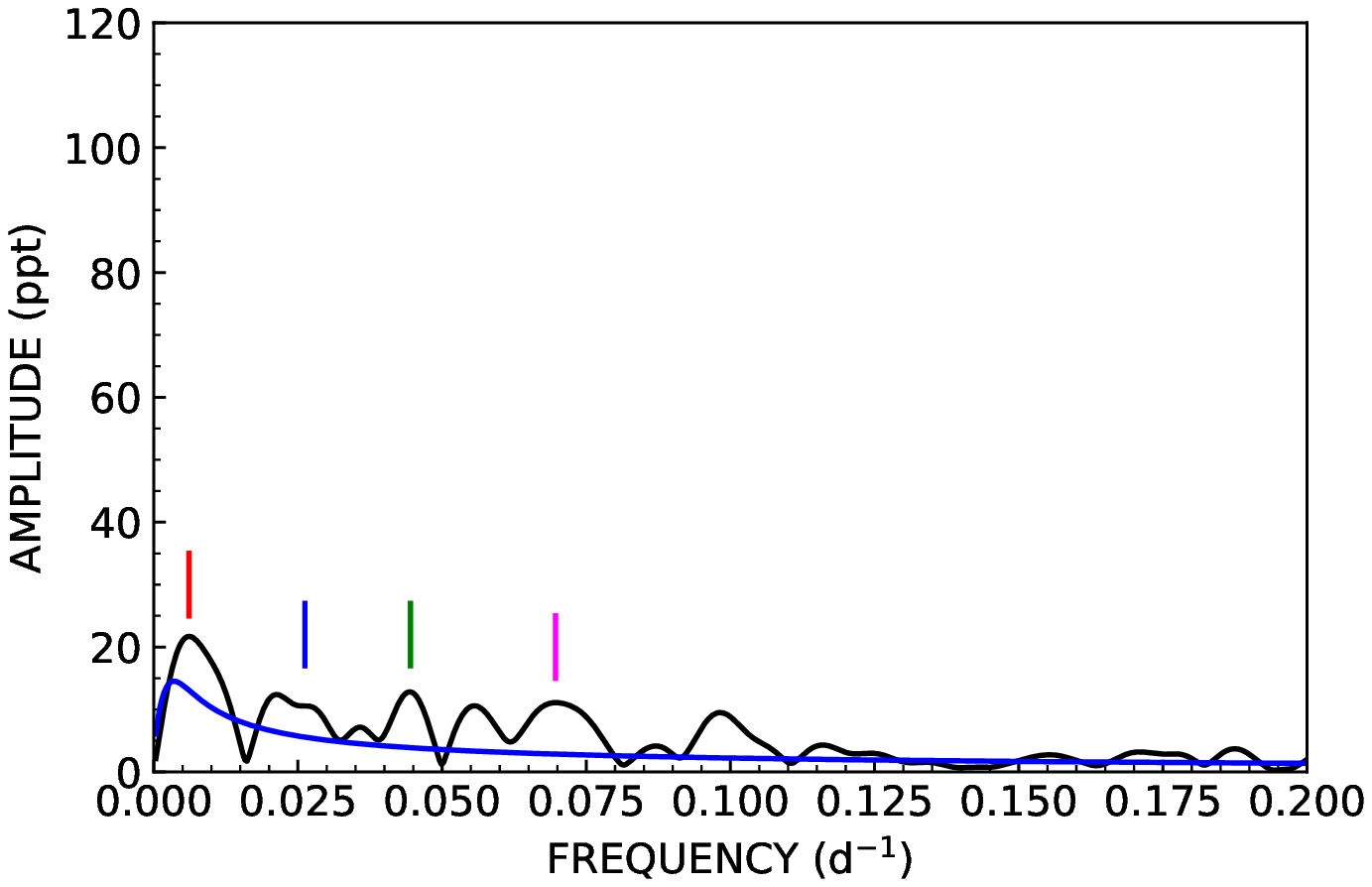}
	\caption{The {\it BRITE} flux, after subtracting off the global mean, with units of parts per thousand (ppt) (left) and the Fourier amplitude spectrum (right) for the 2016 data from {\it BRITE}, format as in Fig. \ref{fig:2014}. }
    \label{fig:2016}
\end{figure*}

\begin{figure*}
    \includegraphics[width=\columnwidth]{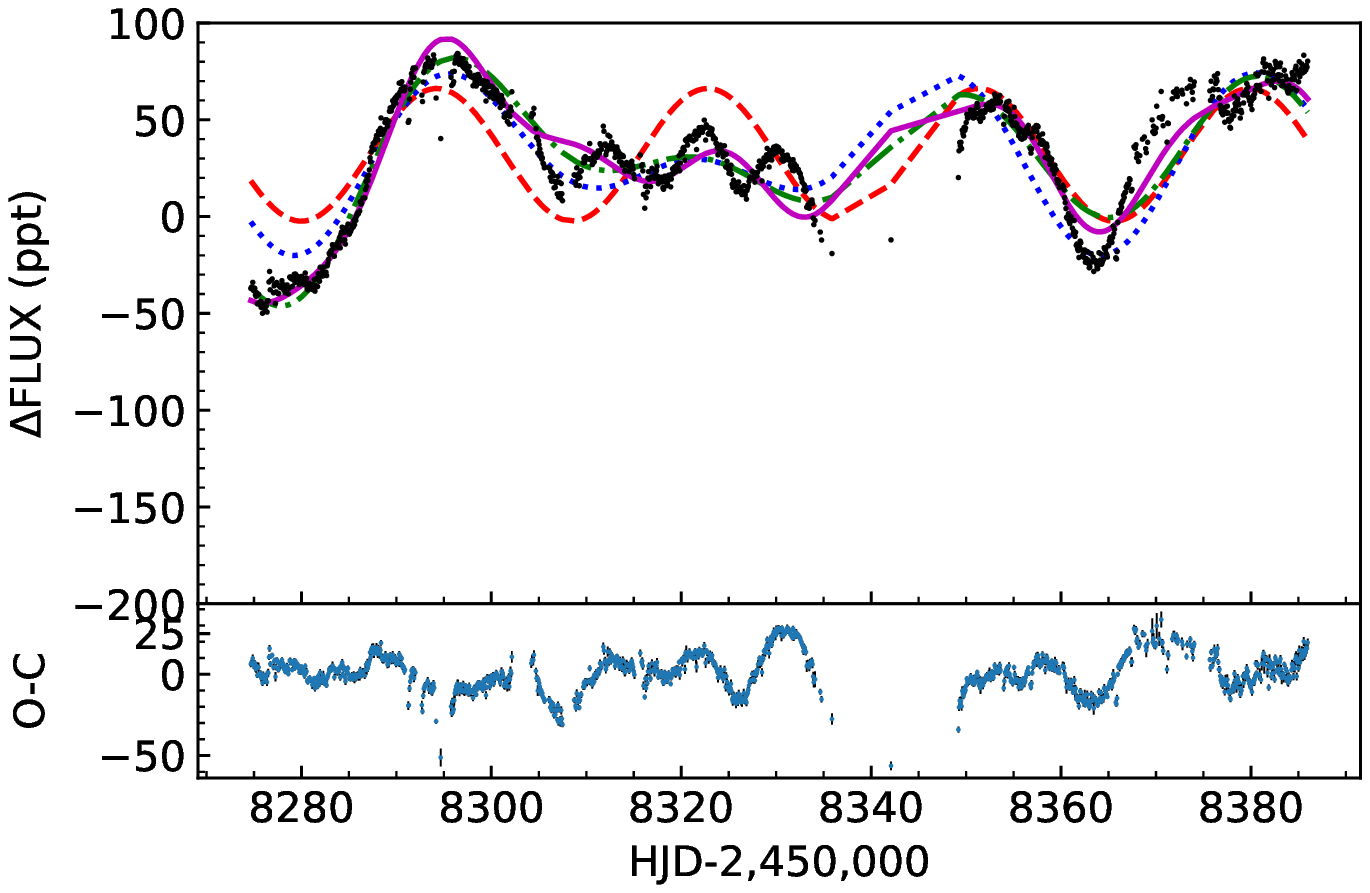}
    \includegraphics[width = \columnwidth]{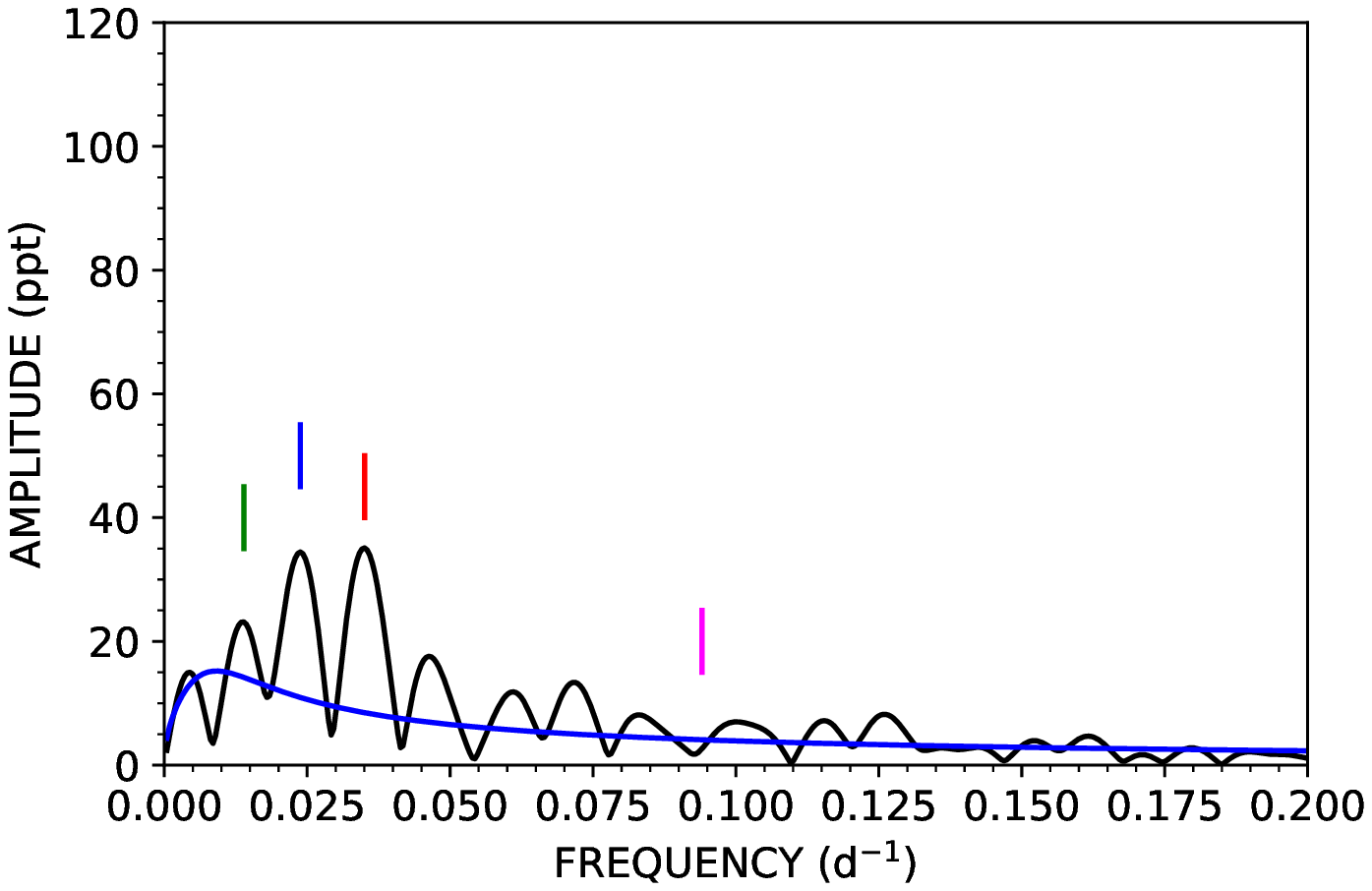}
	\caption{The {\it BRITE} flux, after subtracting off the global mean, with units of parts per thousand (ppt) (left) and the Fourier amplitude spectrum (right) for the 2018 data from {\it BRITE}, format as in Fig. \ref{fig:2014}. }
    \label{fig:2018}
\end{figure*}

\begin{figure*}
    \includegraphics[width=\columnwidth]{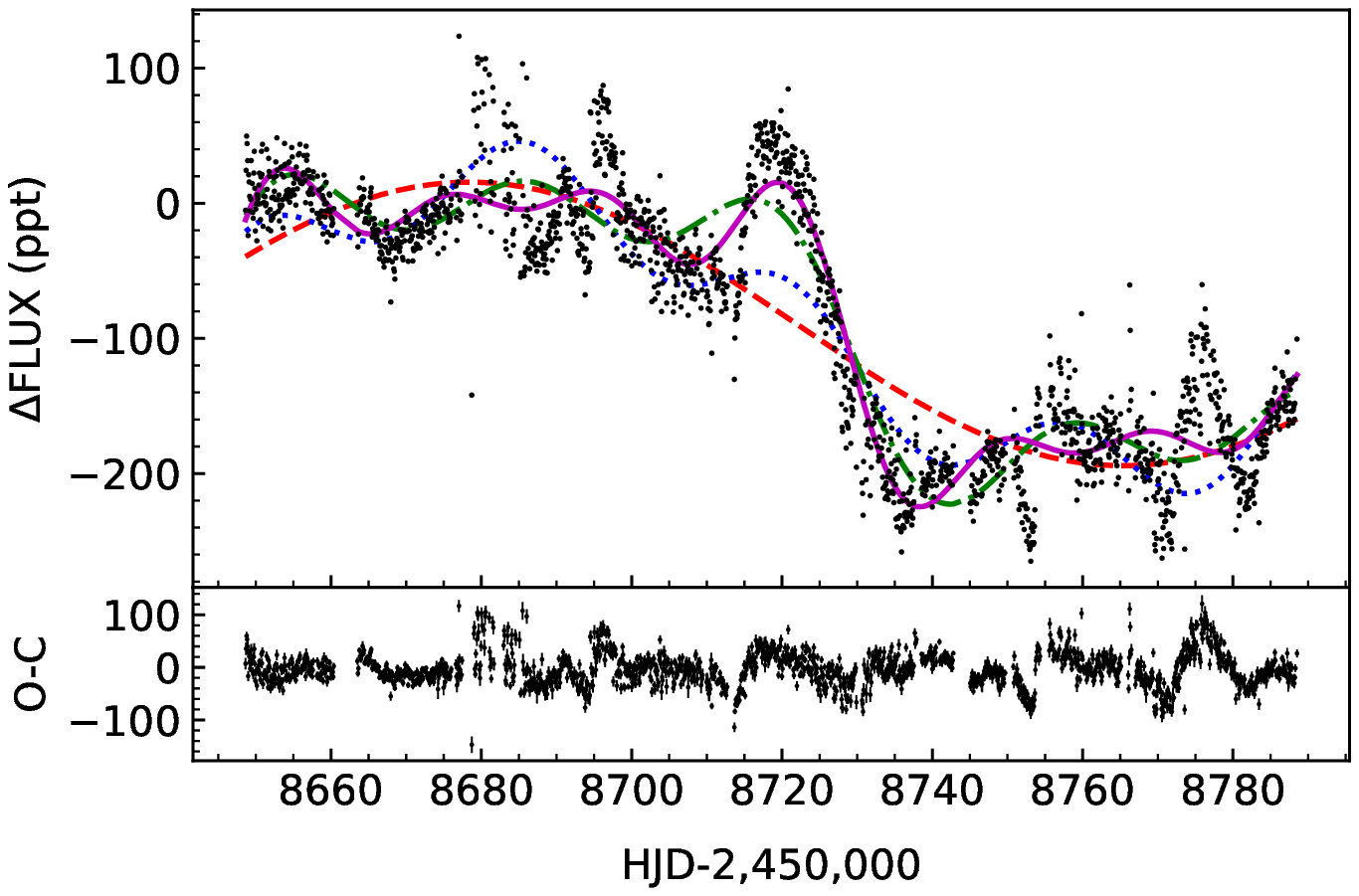}
    \includegraphics[width = \columnwidth]{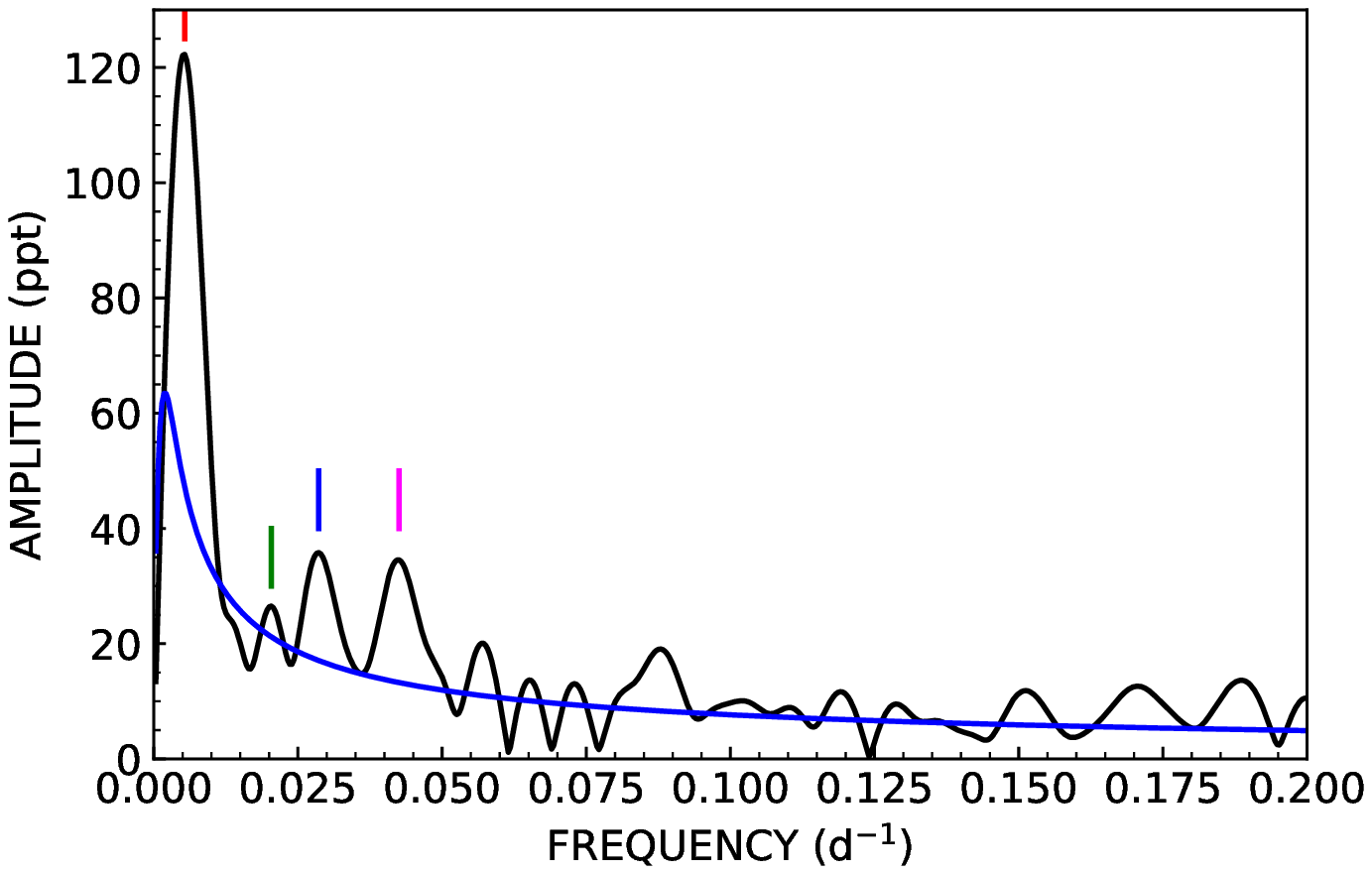}
	\caption{The {\it BRITE} flux, after subtracting off the global mean, with units of parts per thousand (ppt) (left) and the Fourier amplitude spectrum (right) for the 2019 data from {\it BRITE}, format as in Fig. \ref{fig:2014}. }
    \label{fig:2019}
\end{figure*}

\section{Fourier Analysis}

To find significant periodicities in the {\it BRITE} data, we performed a Fourier analysis. To begin, we examined our complete light curve (Fig. \ref{fig: all data}), and then calculated the Fourier spectrum with the {\tt Period04} software available from \citet{2005CoAst.146...53L}. With this analysis, we found one peak at a frequency of 0.001045 d$^{-1}$, corresponding to a period of 956.8 d (2.65 years). We utilized the methods described by \citet{2017MNRAS.467.2494P} to determine the noise limit of these data through a false alarm probability and then calculate a signal-to-noise ratio based on the strength of the peak divided by the noise level at that frequency. We also calculate a $\sigma(O-C)$ as the standard deviation of the quantity of the difference between the observed and calculated points, the $(O-C)$ curve, after that fit. If adding additional terms from the Fourier analysis does not alter the value of $\sigma(O-C)$, then additional terms can be rejected.
Our full analysis of the light curve shows that the 956.8 d ``period" was marginally detected with a signal-to-noise ratio of 1.9 and an amplitude of 63.4 parts per thousand (ppt), as shown in Fig.~\ref{fig: all data}. No additional frequencies significantly improved the fit to the full {\it BRITE} data set. 

\begin{table*}
\begin{tabular}{llllll}
Term number	&	Frequency ($d^{-1}$)			&	Amplitude (ppt)			&	Phase			&	$\sigma_{\rm(O-C)}$ (ppt)	&	S/N	\\ \hline
																
\multicolumn{6}{c}{2014, ZP: -0.02556	$\pm$	0.00043}		\\ \hline
																	
1	&	0.02553	$\pm$0.00029	&	14.45	$\pm$0.77	&	0.02498	$\pm$0.067	&	19.13	&	1.39	\\
																	
2	&	0.08469	$\pm$0.00026	&	16.89	$\pm$0.72	&	0.6762	$\pm$0.0060	&	15.15	&	3.02	\\
																	
3	&   0.01375	$\pm$0.00024	&	19.39	$\pm$0.66	&	0.1637	$\pm$0.0057	&	13.33	&	1.60	\\
																	
4	&	0.05780	$\pm$0.00029	&	14.31	$\pm$0.58	&	0.7383	$\pm$0.0070	&	10.70	&	2.06	\\ \hline

\multicolumn{6}{c}{2015, ZP:  0.00019	$\pm$	0.00029} \\ \hline
																	
1	&	0.00428	$\pm$	0.00004	&	55.26	$\pm$0.36	&	0.7778	$\pm$	0.0012	&	17.92	&	3.56	\\
																	
2	&	0.02966	$\pm$	0.000088	&	14.51	$\pm$0.32	&	0.0844	$\pm$	0.0037	&	14.78	&	2.98	\\
																	
3	&	0.05474	$\pm$	0.00010	&	12.95	$\pm$0.35	&	0.8960	$\pm$	0.0038	&	12.45	&	3.78	\\
																	
4	&	0.07937	$\pm$	0.00016	&	7.957	$\pm$0.34	&	0.5445	$\pm$	0.0062	&	10.81	&	2.92	\\ \hline

	\multicolumn{6}{c}{	2016, ZP:  -0.0125	$\pm$	0.0016}	\\ \hline
																	
1	&	0.00554	$\pm$	0.00062	&	27.34	$\pm$	1.32	&	0.2563	$\pm$	0.0028	&	20.91	&	2.27	\\
																	
2	&	0.02482	$\pm$	0.00018	&	14.99	$\pm$	0.93	&	0.8460	$\pm$	0.0047	&	18.15	&	2.91	\\
																	
3	&	0.04487	$\pm$	0.00020	&	13.10	$\pm$	0.69	&	0.4488	$\pm$	0.0056	&	15.28	&	2.98	\\
																	
4	&	0.07183	$\pm$	0.00030	&	11.73	$\pm$	4.14 &	0.9263	$\pm$	0.0062	&	12.98	&	3.86	\\ \hline
\multicolumn{6}{c}{2018, ZP: 0.03146	$\pm$	0.00042	}	\\ \hline 
																	
1	&	0.03246	$\pm$	0.0062	&	41.70	$\pm$	1.32	&	0.2613	$\pm$	0.0034	&	22.81	&	4.66	\\
																	
2	&	0.02675	$\pm$	0.0002	&	41.75	$\pm$	0.92	&	0.4738	$\pm$	0.0032	&	15.61	&	4.09	\\
																	
3	&	0.01989	$\pm$	0.0002	&	18.62	$\pm$	0.69	&	0.1041	$\pm$	0.0040	&	12.45	&	1.53	\\
																	
4	&	0.09407	$\pm$	0.0030	&	7.766	$\pm$	4.14	&	0.4059	$\pm$	0.021	&	11.19	&	1.90	\\ \hline 
																	
\multicolumn{6}{c}{2019, ZP: -0.01247	$\pm$	0.0016}	\\ \hline
																	
1	&	0.00631	$\pm$	0.0062	&	109.3	$\pm$	1.32	&	0.06254	$\pm$	0.0017	&	47.85	&	2.55	\\
																	
2	&	0.02934	$\pm$	0.00018	&	35.80	$\pm$	0.92	&	0.9423	$\pm$	0.0049	&	42.57	&	2.13	\\
																	
3	&	0.01784	$\pm$	0.00020	&	35.76	$\pm$	0.69	&	0.3644	$\pm$	0.0055	&	35.22	&	1.55	\\
																	
4	&	0.04158	$\pm$	0.00030	&	23.91	$\pm$	4.14	&	0.8699	$\pm$	0.0070	&	31.37	&	1.94	\\ \hline
\multicolumn{6}{c}{All Data, ZP: -0.03127	$\pm$	0.00067	}	\\ \hline
																	
1	&	0.00105	$\pm$	0.0023	&	63.43	$\pm$	0.87	&	0.09368	$\pm $	0.0030	&	53.12	&	1.95	\\
\hline

\end{tabular}
\caption{\label{FTTable} The resulting coefficients from the Fourier analysis of our data. The phase is relative to the zero-point of the HJD calendar.}
\end{table*}

Next we analysed individual {observational campaigns} of {\it BRITE} observations. For each season of {\it BRITE} photometry, we performed a similar analysis using the {\tt Period04} software and found four frequencies that we include in our analysis. These fits are shown in Figures \ref{fig:2014}--\ref{fig:2019}. For each season, we present in Table \ref{FTTable} the term number $i$ that we use in the equation 
$$ F = ZP + \sum_{i=1}^{4} A_i \sin{(2\pi(f_i t + \phi_i))} $$
to reproduce the light curve, where $ZP$ is a zero point of the data, $f_i$ is the $i^{\rm th}$ frequency and $\phi_i$ is a phase. These are all relative to time $t$, which was calculated to be relative to the heliocentric Julian Date (-2,450,000). Each figure for the datasets (Fig.~\ref{fig:2014}-\ref{fig:2019}) shows the addition of the individual terms as additional sine waves with the different colors of the associated sine waves. Each iteration was then improved upon so that some peaks we derive do not appear directly in the original Fourier transform (e.g., Fig.~\ref{fig:2018}), but this is in part due to the low signal-to-noise of these peaks.

The signal-to-noise of the derived peaks is typically low, with a S/N usually being $\lesssim 2$. \citet{2015MNRAS.448L..16B} showed that for the more precise, higher cadence $K2$ data, a S/N of $5$ is preferred as to not over interpret the light curve with respect to pulsations. We also measure a goodness of fit with a standard deviation of the residuals of the $(O-C)$ curve, which we report as $\sigma_{\rm (O-C)}$ in Table \ref{FTTable}.

\section{Discussion}

{We have collected the first high-precision and high-cadence light curve of a luminous blue variable.} This has allowed us to explore the light curve with Fourier analysis in an attempt to discover periodicities that are intrinsic to the star. 

Our Fourier analysis presented in Section 3, did not produce any frequencies that were significant with a S/N of 4 or higher based on the methods described by \citet{2017MNRAS.467.2494P}. Since the light curve of this typical luminous blue variable is most certainly variable compared to the errors of the data, we must therefore look for alternate explanations of the variability.

In a recent spectroscopic analysis, 
\citet{2020JAVSO..48..133P} reported a periodic behavior of the H$\alpha$ equivalent width for P Cygni, and reported a 318-d period for the time period covering 2005--2019. If confirmed, a strict periodicity would be extremely interesting as no other study has found a strict periodicity for P Cygni. In contrast, \citet{2018MNRAS.475.5417R} found that there are two periods in the {\it BRITE} light curves of $\eta$ Carinae over two years, which they were able to compare to the tidally excited oscillations that have been seen for many eccentric binaries across the H-R diagram. 

P Cygni is not a well-established binary. \citet{2010MNRAS.405.1924K} modelled the eruptive light curve as caused by binarity, but the light curve from the time of the eruption is not a well-recorded time-series. The light curve modelled by \citet{2010MNRAS.405.1924K} was actually that of \citet{1988IrAJ...18..163D}, which
was re-evaluated by \citet{2011MNRAS.415..773S} who concluded that the eruptive light curve of P Cygni has a more “typical” appearance of an LBV eruption when we consider the sparsely sampled light curve. \citet{2013ApJ...769..118R} found no indication of a companion from interferometric observations with a limiting magnitude difference of $\Delta H$ = 5.3, while a recent examination of the multiplicity of LBVs by \citet{2021arXiv210512380M} reported a possible companion with $\Delta H$ = 4.3, a full magnitude brighter than the limiting magnitude reported by \citet{2013ApJ...769..118R}, although \citet{2021arXiv210512380M} searched over a larger field-of-view.

\begin{figure}
	\includegraphics[width=\columnwidth]{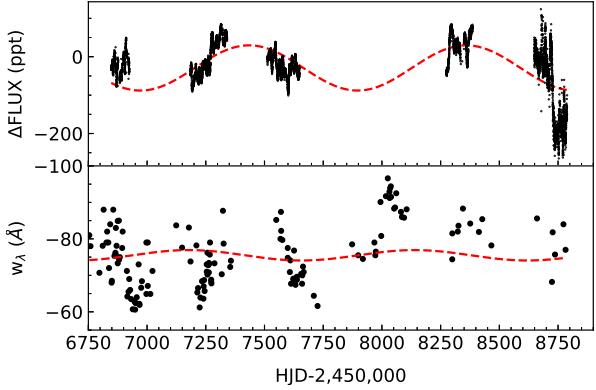}
    \caption{We compare our results on the Fourier analysis of all of the {\it BRITE} data to the measured H$\alpha$ equivalent widths measured by \citet{2020JAVSO..48..133P}. {The 950-d period from our photometric analysis (see Fig. 3) is shown as a dashed red line, and is also forced to fit the H$\alpha$ measurements.} }
    \label{fig:pollmann}
\end{figure}

If the 318 d period reported by \citet{2020JAVSO..48..133P} is confirmed, it could represent a pulsation mode driven by tidal forces, such as that in the massive binary $\iota$ Ori \citep{2017MNRAS.467.2494P} or the LBV binary $\eta$ Car \citep{2018MNRAS.475.5417R}. {This could also represent a normal pulsation for supergiants \citep{1980A&A....90..311M}}. However, we cannot photometrically confirm the 318 d period that was reported from the study of H$\alpha$ equivalent widths, {although it is possible that the $318$ d period shown in Fig.~\ref{fig:pollmann} could represent a harmonic of the $\sim950$ d period. Unfortunately, the significance of that periodicity in the {\it BRITE} data is only 2$\sigma$ above the noise level of the Fourier spectrum (Table \ref{FTTable}). } We compare these data sets in Fig.~\ref{fig:pollmann}, but also caution that searching for a 318-d period in our {\it BRITE} data is difficult owing to the length of the data-sets ($\sim 150$~d) and their being spaced by $\sim 0.5$ yr apart.

Both \citet{2001A&A...366..935M, 2001A&A...376..898M} and \citet{2011AJ....141..120R} studied the long-term photometric trends in relation to the trends observed with H$\alpha$. These studies all found a positive correlation between the H$\alpha$ equivalent widths and $V$-band photometry. We calculated the needed amplitude and phase for a period that was the same as our derived photometric period shown in Fig.~\ref{fig: all data}, and show these in Fig.~\ref{fig:pollmann}. 

When comparing the H$\alpha$ measurements over the same time period as the photometry, we found that the two data sets could show similar variability time scales. The raw curves show similar trends, with an increase in flux and equivalent width across the 2015 data set and a decrease in 2016. The 2018 and 2019 H$\alpha$ data sets were too sparse to see similar trends in the data. However, the times of the maxima derived from this analysis are different for both data sets. The data-sets are likely correlated, but the gap in the {\it BRITE} observations in 2017 and the sparse ground-based observations of H$\alpha$ will prevent a firm analysis of this correlation, and any photometric periodicity lies below a S/N of 4 that is typical for confirmation of periodicities. 
 
 \subsection{Properties of the stochastic low-frequency variability}
 
Recent advances in astronomical time-series have revolutionized our understanding of massive star asteroseismology \citep[see][for a recent review]{Bowman2020c}. With the advent of all-sky surveys searching for planet transits, time-series photometry of massive stars has become a useful tool in studying these types of stars, and two general explanations have been used to describe the typical properties of the Fourier transforms for massive stars, namely internal gravity waves \citep{Rogers2013b, Edelmann2019a,Horst2020a} and sub-surface convection \citep{Lecoanet2019a, Cantiello2021b}.
P Cygni is a {well-established} LBV, so the long time-series of {\it BRITE} photometry can be considered a first step towards understanding the driving mechanisms for the long (i.e. $\gtrsim 100$~d) and short time-scale (i.e. $\lesssim$~d) variability of luminous blue variables. We have compiled a table of reported time-scales and ``periods" for P Cygni in Table \ref{timescales}, along with the interpretation or discovery type of these variations.

\begin{table}
\begin{tabular}{lll}
Time scale & Interpretation & Reference \\
    & or Discovery type &   \\
\hline
$\sim$ 7.3 yr & SD-phase & 1 \\
1700 d      & H$\alpha$ discrete absorption component   & 2 \\
    &    recurrence time & \\
$\sim$ 4 yr & SD-phase &   3, 4, 5, 6 \\
           & radial pulsations & 3, 5\\
           & gravity wave, but non-radial pulsations & 7 \\
 $\sim 900-1000$ d    &   long-term photometry  & This work \\   
    &   gravity waves, sub-surface convection & \\
 $\sim 700$ d    &   H$\alpha$ discrete absorption component  &   2 \\    
    &   progression time & \\
 $\sim$ 500 d & pressure wave & 7 \\
 318.3 d & H$\alpha$ equivalent width & 8\\
 $\sim$175 d & pressure wave & 7 \\
 $\sim$100 d & pressure wave or other type of oscillation & 4, 5, 7 \\
17.3 d      & $\alpha$ Cyg osc. & 4, 5 \\
\hline
\end{tabular}
\caption{\label{timescales} Previous periods and timescales reported for P Cygni with their interpretation and discovery method. References are 
1: \citet{2000ASPC..204..111M}, 
2: \citet{2011AJ....141..120R}, 
3: \citet{2001A&A...366..935M}, 
4: \citet{2001ASPC..233...15D}, 
5: \citet{2001A&A...376..224D},
6: \citet{2001A&A...376..898M},
7: \citet{2001ASPC..233..215D},
and 8: \citet{2020JAVSO..48..133P}.
}
\end{table}

The time-scales in Table \ref{timescales} show P Cygni is known to be variable with periods between 17 d and several years. These have been interpreted as $\alpha$ Cygni-like oscillations for the shortest period oscillations \citep[e.g.,][]{2001A&A...376..224D}, consistent with the ideas that the luminous blue variables are extreme versions of normal luminous supergiants. In Figures \ref{fig:2014}--\ref{fig:2019}, the light curves of P Cygni throughout the {\it BRITE} campaign are shown. Short-term variations are certainly seen, but rarely is it similar to a 17-d timescale, and that particular period is never required to fit the light curves in Figs.~\ref{fig:2014}--\ref{fig:2019}.

\begin{figure}
    \centering
    \includegraphics[width = 3.5in]{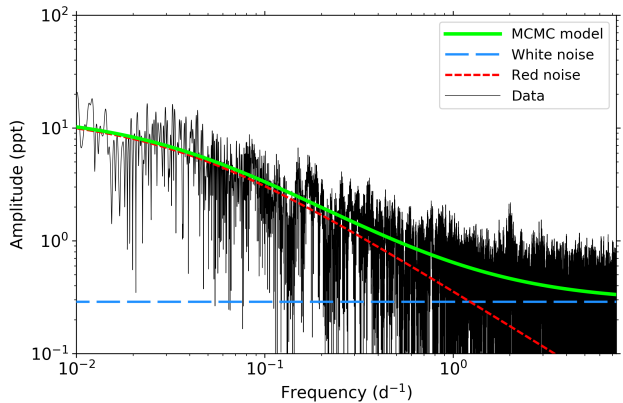}
    \caption{The white- and red-noise fit to the Fourier transform of the {\it BRITE} data. The model is described in the text. }
    \label{bowman}
\end{figure}

The Fourier transform of the {\it BRITE} data shows that the variability extends over a broad frequency range, and dominated by longer periods. We used the methods of \citet{2019A&A...621A.135B,2019NatAs...3..760B,2020A&A...640A..36B} to fit the white noise and stochastic low-frequency variability components of the Fourier transform with the equation
$$ \alpha \left( \nu \right) = \frac{ \alpha_{0} } { 1 + \left( \frac{\nu}{\nu_{\rm char}} \right)^{\gamma}} + C_{\rm w}. $$
In this equation, $\alpha_{0}$ represents the amplitude at a frequency of zero, $\gamma$ is the logarithmic amplitude gradient, $\nu_{\rm char}$ is the characteristic frequency, which is the inverse of the characteristic timescale, $\tau$, of stochastic variability present in the light curve such that $\nu_{\rm char} = (2\pi\tau)^{-1}$, and $C_{\rm w}$ is a frequency-independent (white) noise term \citep{2011A&A...533A...4B, 2019A&A...621A.135B}. To fit the equation and derive errors, we utilize the python package {\tt emcee} \citep{2013PASP..125..306F} and derive the terms of 
$\alpha_{0} = 12.8\pm0.2$ parts per thousand in fractional flux,
$\nu_{\rm char} = 0.033\pm0.001$ d$^{-1}$,
$\gamma = 1.04\pm0.01$, and
$C_{\rm w} = 0.288\pm0.004$ parts per thousand in fractional flux. While we did not fit each individual observational campaign in this same manner, we note that the Fourier analyses of these data sets, shown in Figs.~\ref{fig:2014}--\ref{fig:2019}, all have their ``peaks" at levels close to the false-alarm probability noise level, with no periodicity seen with a S/N $\gtrsim$ 4. Given the small value of $\nu_{\rm char}$, we note that individual data sets may not be of sufficient duration to fit this for every year of data. The resultant fit of the Fourier transform of the {\it BRITE} data up to the Nyquist frequency is shown in Fig.~\ref{bowman}, which shows that significant variability above the white noise level extends up to $\nu \simeq 1$~d$^{-1}$.

The results of \citet{2019NatAs...3..760B, 2020A&A...640A..36B} have been used to interpret the variability of {\it K2} and {\it TESS} light curves of massive stars as the result of internal gravity waves. Their sample was composed primarily of O and B stars on or near the main sequence. A comparison to the H-R diagram in \citet[][their Fig.~2]{2020A&A...640A..36B}, we can see that while the effective temperatures of their sample are similar to that of LBVs, most of the stars they analyzed have a luminosity an order of magnitude smaller than P Cygni \citep[$\log L/L_\odot$ = 5.85; $\log T/K$ = 4.28; ][]{1997A&A...326.1117N}. Therefore, given that \citet{2020A&A...640A..36B} focused on a sample of main sequence stars using TESS data, a direct quantitative comparison with our analysis of P Cygni using BRITE data is not possible. Yet, the trends in the fits shown in \citet{2020A&A...640A..36B} shows that the characteristic frequency, $\nu_{\rm char}$, decreases with increasing luminosity and decreasing temperature, while the amplitude, $\alpha_{0}$, increases  with increasing luminosity and decreasing temperature. P Cygni, representing a star with a luminosity an order of magnitude higher than these other blue supergiants fits this trend with a characteristic frequency almost an order of magnitude lower and an amplitude an order of magnitude higher than the stars in the sample of \citet{2020A&A...640A..36B}. 

\citet{2011AJ....141..120R} found that there was no preferred period or timescale in a long 24-year time-series of either spectroscopic measurements or photometry. Their Fourier analysis showed an increase in power with lower frequency, with a shape similar to that shown in Fig.~\ref{bowman}. The nature of these variations and timescales show that the star is not a stable pulsator, but does have stochastic low-frequency variability. In many ways, P Cygni seems to behave similarly to the OB stars analyzed by \citet{2020A&A...640A..36B}. 
In addition to internal gravity waves as a dominant driver of the stochastic low-frequency variability, sub-surface convection could also be responsible. However, the properties of convection and its observational consequences in terms of predicting the time scales of stochastic low-frequency variability in light curves within the parameter space of P Cygni have yet to be fully explored both theoretically and using numerical simulations \citep[e.g.,][]{Jiang_Y_2015}. Furthermore, as stars leave the main sequence, stochastic low-frequency variability caused by stellar winds is also expected to become more important, especially in more evolved stars \citet[e.g.,][]{Aerts2018a, Krticka2021b}. Our observational study will help guide future theoretical studies of these phenomena. 
With the {\it TESS} mission regularly observing the Large Magellanic Cloud, a larger sample of LBVs may be able to be fit similarly in the future for studying the driving mechanisms of these stars as a population.

\section{Conclusions}

In this paper, we have studied the luminous blue variable P Cygni with a long-time series of precision photometry collected with the {\it BRITE-Constellation} of nanosatellites. We have found some interesting results with these data:
\begin{itemize}
    \item The analysis of the {\it BRITE} data on P Cygni indicates that the main driver of the variability of this luminous blue variables is similar in morphology in the Fourier spectrum as internal gravity waves seen in main sequence OB stars \citep{2020A&A...640A..36B} and stellar winds in evolved stars \citep{Krticka2021b}. Since no single frequency is persistent across these data sets, the variability is stochastic and not strictly periodic, as has been discussed in past studies.
    \item Each season of {\it BRITE} data could be modelled with four frequencies, but these frequencies are not of high significance.
    \item There is some evidence of a correlation between H$\alpha$ variations and the flux as measured by {\it BRITE} (Fig.~\ref{fig:pollmann}).
    \item There is no single frequency that is persistent in all of these data, and the previously reported $\sim17$ day period seen in many datasets \citep[e.g.,][]{2001A&A...376..224D} is not seen in these photometric data at any time. Unlike the massive LBV binary $\eta$ Car, this lack of persistent period may be an indication that this LBV is a single star showing only stochastic variability, unlike the tidally excited oscillations seen in $\eta$ Carinae. 
\end{itemize}

\section*{Acknowledgements}

We thank the anonymous referee for comments that improved the presentation of this paper. We acknowledge with thanks the variable star observations from the AAVSO International Database contributed by observers worldwide and used in this research, and in particular the necessary reductions of the {\it BRITE} photometry. We also acknowledge Ernst Pollmann for making the H$\alpha$ measurements available to us. Adam Popowicz was responsible for image processing and automation of photometric routines for the data registered by {\it BRITE}-nanosatellite constellation and was supported by grants: 02/140/RGJ21/0012 (SUTO Rector’s grant) and BK-225/RAu-11/2021 (Statutory Activities Grant). AFJM is grateful for financial assistance from NSERC (Canada). GH thanks the Polish National Center for Science (NCN) for support through grant 2015/18/A/ST9/00578. NSL wishes to thank the Natural Sciences and Engineering Council (NSERC) of Canada for financial support. DMB gratefully acknowledges a senior postdoctoral fellowship from the Research Foundation Flanders (FWO) with grant agreement no.~1286521N. GAW acknowledges Discovery Grant support from the Natural Sciences and Engineering Council (NSERC) of Canada.

\section*{Data Availability}

The raw {\it BRITE} data are available on the {\it BRITE} Public archive.\footnote{https://brite.camk.edu.pl/pub/index.html}. The AAVSO photometry used in the reductions are available at the AAVSO database \footnote{https://www.aavso.org/}. Our reduced light curve is available as a machine-readable table with this publication.

\bibliographystyle{mnras}
\bibliography{example}

\appendix

\section{The full BRITE and AAVSO light curve}
\label{sec:appendix}

\begin{figure*}
    \centering
    \includegraphics[width=5.5in]{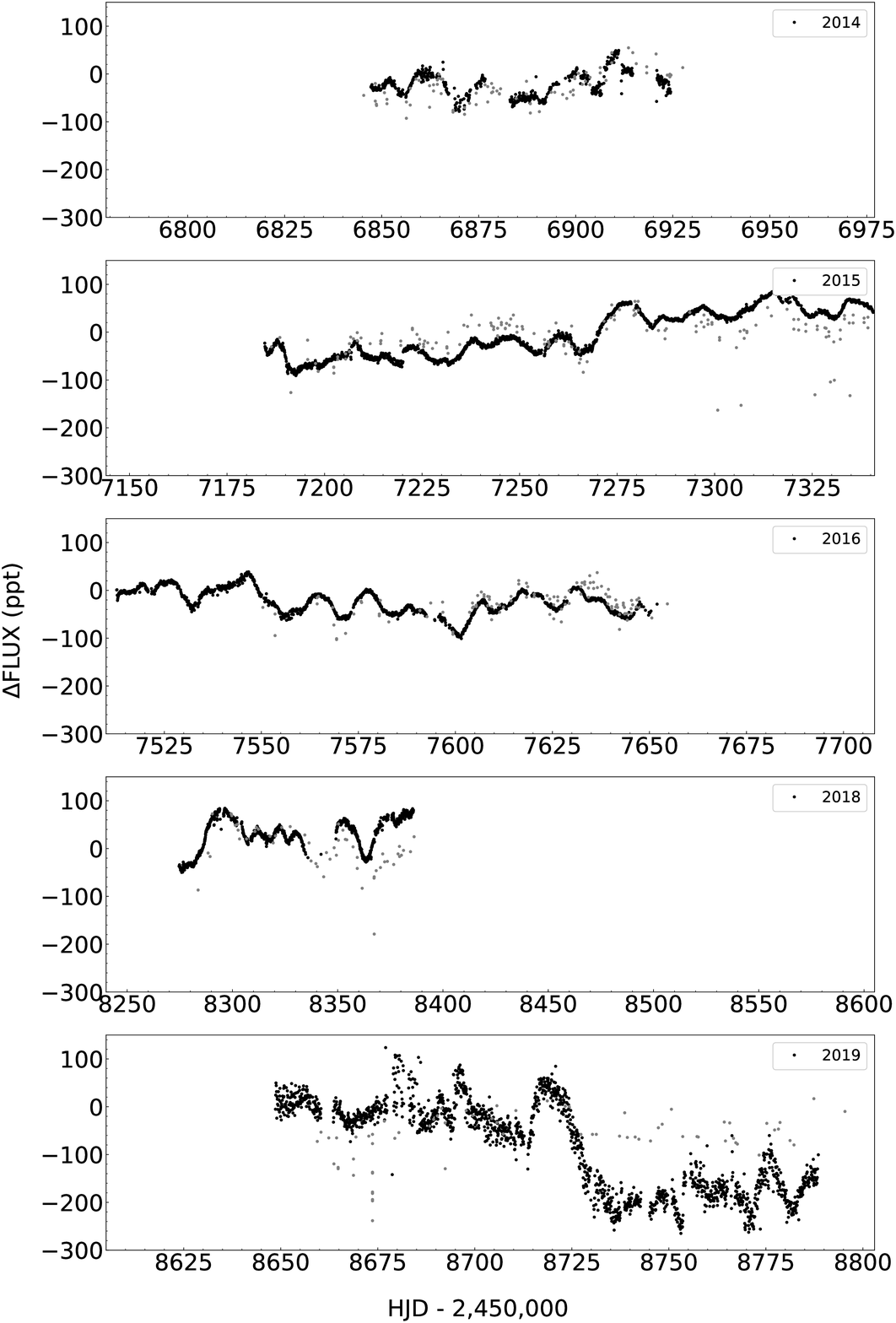}
    \caption{The full BRITE light curve after reductions, with the x-range for each sub-panel representing the time frame of May 15--November 30 of each observing year. {In small grey points, we show the AAVSO light curve, converted to differential flux, for comparison.} }
    \label{fig:appendixcompare}
\end{figure*}

\bsp	
\label{lastpage}
\end{document}